# Open diffusion MRI and connectivity data for epilepsy and surgery: The IDEAS II release

Peter N. Taylor[1,2,3]*, Gerard Hall[1], Jonathan Horsley[1], Yujiang Wang[1,2,3], Sjoerd B. Vos[4,5], Gavin P Winston[3,6], Andrew W McEvoy[3], Anna Miserocchi[3], Jane de Tisi[3], John S Duncan[3]

1. CNNP Lab (www.cnnp-lab.com), Interdisciplinary Computing and Complex BioSystems Group, School of Computing, Newcastle University, Newcastle upon Tyne, United Kingdom

2. Faculty of Medical Sciences, Newcastle University, Newcastle upon Tyne, United Kingdom

3. UCL Queen Square Institute of Neurology, Queen Square, London, United Kingdom

4. Centre for Medical Image Computing, Department of Computer Science, UCL, London, United Kingdom

5. Centre for Microscopy, Characterisation, and Analysis, The University of Western Australia, Nedlands, Australia

6. Department of Medicine, Division of Neurology, Queen's University, Kingston, Canada

* Peter.Taylor@newcastle.ac.uk

# Abstract

Epileptic seizures are generated in cerebral networks that propagate ictal and interictal activity. The structure of cerebral networks underpinning epileptic activity can be inferred from diffusion-weighted MRI (DWI). However, publicly available DWI data in individuals with epilepsy are scarce, and processing is technically challenging due to scan-specific artifacts, limiting research progress.

Here, we release raw DWI data from 216 individuals with epilepsy and 98 healthy controls. Subject identifiers align with our previous data release (IDEAS), which includes T1-weighted and FLAIR MRI, surgical details, and long-term seizure outcomes after surgery. Preprocessing reduced distortions and artifacts, while fully processed data include diffusion metric maps in native and template space. We also provide parcellated structural connectomes using multiple atlases and connectivity measures.

To illustrate the utility of this IDEAS II data, we replicated ENIGMA consortium findings, observing widespread reductions of fractional anisotropy, particularly ipsilateral to the area of seizure onset. We further demonstrate localised abnormality, and network connectivity using streamline tractography in a patient who subsequently underwent temporal lobe resection.

This open dataset offers a comprehensive resource to advance research on structural connectivity and surgical outcomes in epilepsy.

# Introduction

Data sharing initiatives such as the human connectome project, ADNI, and ABIDE have facilitated major progress in their respective fields[1–5]. In epilepsy neuroimaging, only very limited public data exist (though see[6] for one example). Previously we shared the largest epilepsy neuroimaging data ever made publicly available - the Imaging Database for Epilepsy And Surgery (IDEAS,[7]). That database contains T1-weighted and FLAIR MRI acquisitions, additional to resection masks generated from post-operative imaging. Rich clinical/demographic metadata are also included, and IDEAS has already facilitated several further studies[8–10].

Though valuable, the IDEAS data in its original release does not allow the inference of white matter structural networks or microstructural properties. This is an important limitation, since several studies demonstrate white matter alterations in epilepsy[11], their importance for planning surgical treatment and mitigating post-operative sequelae[12–15], and complementarity to neurophysiology when localizing epileptic activity[16].

Diffusion weighted MRI (DWI) allows the inference of microstructural properties and white matter networks. Furthermore, DWI is recommended in epilepsy during pre-surgical evaluation to identify key tracts for surgical avoidance[17,18]. However, acquisition protocols vary widely, and DWI can be prone to technical artifacts which, if not addressed, can lead to major problems.

In this work, we build on our previous data release by sharing data from 216 people with drug resistant focal epilepsy and 98 healthy controls. We term this the IDEAS II data. All DWI scans were acquired using one of two acquisition protocols, with healthy controls acquired with each protocol. Fully preprocessed data are also available, as are structural connectivity matrices.

## Methods

### Study approval

This study of anonymized data that had been acquired previously was approved by the Health Research Authority, without the necessity to obtain individual subject consent (UCLH epilepsy surgery database: 22/SC/0016), and by the Database Local Data Monitoring Committee. Individuals who had declined for their data to be used in anonymized research were not included in the research database.

### Patient and data selection

Data released in this paper are from a subset of the same subjects in our previous IDEAS release, the selection criteria for which have been described previously[7]. Included in this release are those subjects for whom DWI data was acquired pre-operatively and passed quality control.

For 83 (95) patients (controls) the DWI data and IDEAS T1w data were acquired during the same session, an example of which ('sub-1') is shown in figure 1. For the remainder, the DWI and T1w scans were acquired in different sessions. In total 123 (3) patients (controls) had DWI before the T1w scan, and 10 (0) with DWI after the T1w scan. When data were derived from multiple scan sessions, the interval was less than 24 months in 69% of individuals, and less than 36 months in 81% of individuals. In IDEASII, if scans were acquired in different sessions, these are denoted 'ses-1' and 'ses-2'. Note that T1w and FLAIR scans shared in IDEAS II are identical to those already shared in IDEAS I.

### Clinical and demographic information

Clinical and demographic information are included in our previous data release and are identical. Released metadata includes age, sex, histopathology, surgical outcomes, and seizure history. Almost all control scans (98 of 100) released previously are shared with DWI in the current release, and includes age and sex information.

### MRI scan information

Scans were acquired using one of two acquisition protocols described previously[19]. Scan details are included in the data release as javascript object notation (JSON) files, and are summarised below.

Protocol 1 data (N=167) were collected between 2014 and 2019 using a 3T GE MR750 scanner, equipped with a body coil for transmission and a 32-channel phased array coil for reception. Standard imaging gradients with a maximum strength of 50 mTm-1 and slew rate 200 Tm-1s-1 were fitted. Diffusion-weighted MRI data were acquired using a single-shot echo planar imaging (EPI) sequence with echo time = 74.1ms and repetition time 7600ms. Sets of 70 contiguous 2-mm-thick axial slices were obtained covering the whole brain. A total of 115 volumes were acquired with 11, 8, 32, and 64 gradient directions at b-values of 0, 300, 700, and 2500/mm2, respectively ($\delta$ = 21.5ms, $\Delta$ = 35.9ms). The field of view was 25.6 × 25.6cm, and the acquisition matrix size was 128 × 128, giving a reconstructed voxel size of 2 × 2 × 2mm.

Protocol 2 data (N=147) were collected between 2009 and 2013 using a 3T GE Signa HDx scanner equipped with an eight-channel phased array coil. Diffusion MRI was collected using a cardiac-triggered single-shot EPI acquisition [TE = 73ms, TR = heart-rate dependent, b-value of 1200s/mm2 ($\delta$ = 21ms, $\Delta$ = 29ms, using maximum gradient strength of 40 mT m-1), 52 directions with 6 b0. Overall, 60 axial slices were collected, each 2.4-mm thick with a 96 × 96 matrix, zero-filled to 128 × 128 giving, 1.875 × 1.875mm in-plane resolution].

## Data preprocessing

DWI data are frequently prone to artifacts such as EPI-induced distortions (see green arrow indications in figure 1), signal drift, and eddy-current induced distortions. To mitigate these, we release a fully preprocessed version of the data in addition to the raw (unprocessed data). These data are released in the 'preprocessed' folder, and are recommended for use by the research community, though advanced users may wish to apply their own preprocessing pipeline.

To perform preprocessing, the diffusion-weighted imaging (DWI) data were denoised[20], corrected for Gibbs ringing artefacts[21], and B1 field bias using the N4 bias field algorithm[22], all preprocessing steps mentioned were performed with MRtrix3 using: 'dwidenoise', 'degibbs' and 'dwibiascorrect'. We then corrected for signal drift[23]. As some DWI scans did not have a reverse phase encoded B0, all scans were corrected using the Synb0-DISCO tool, which generates a synthetic undistorted b0 image from the individual's T1-weighted scan[24]. The output from Synb0-DISCO was than fed into EDDY[25] from the FSL toolbox[26] to correct for motion and eddy current and EPI induced distortions.

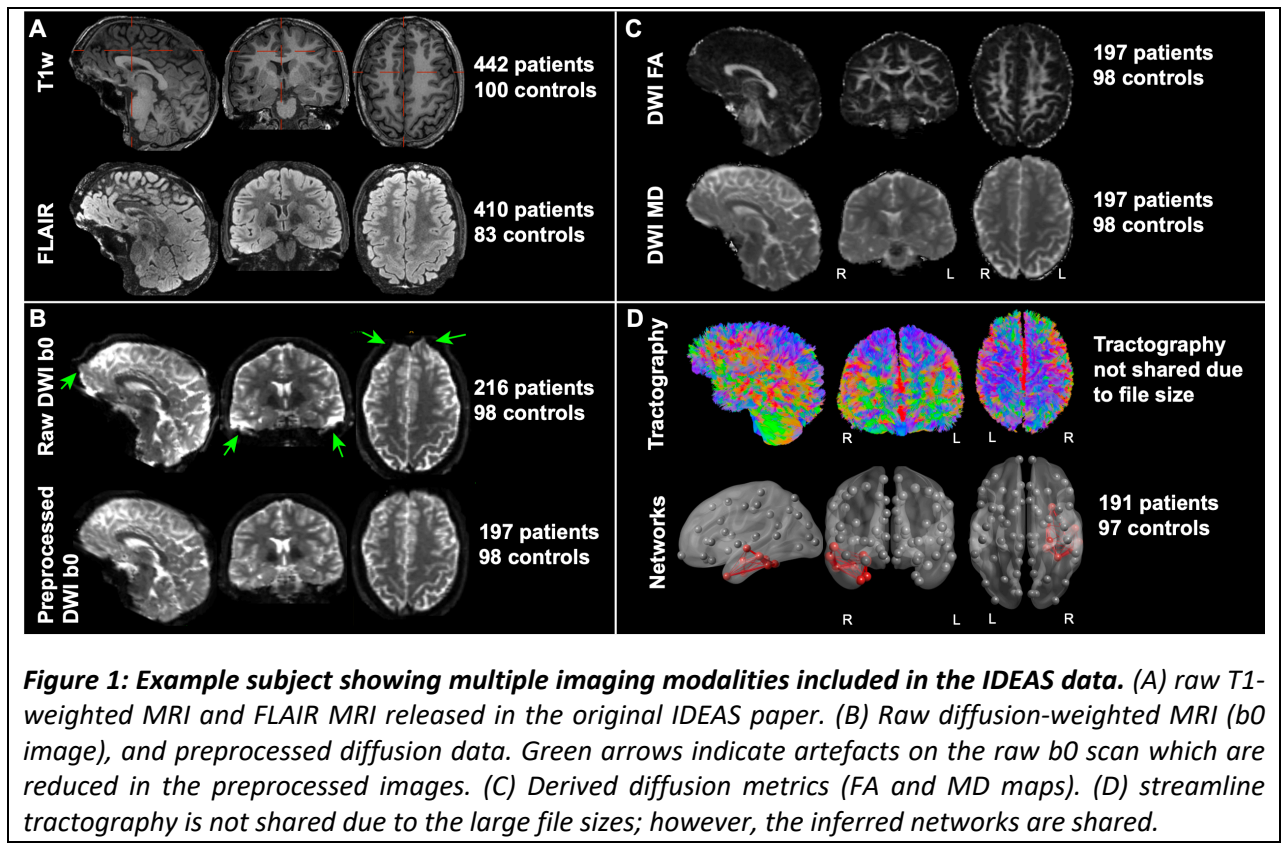

*Figure 1: Example subject showing multiple imaging modalities included in the IDEAS data. (A) raw T1-weighted MRI and FLAIR MRI released in the original IDEAS paper. (B) Raw diffusion-weighted MRI (b0 image), and preprocessed diffusion data. Green arrows indicate artefacts on the raw b0 scan which are reduced in the preprocessed images. (C) Derived diffusion metrics (FA and MD maps). (D) streamline tractography is not shared due to the large file sizes; however, the inferred networks are shared.*

## Data processing

We also share three types of fully processed data. These are shared in the folders 'derivatives/tensor_maps_native', 'derivatives/tensor_maps_MNI152', and 'derivatives/networks'. All use the preprocessed data described in the previous section.

**Tensor maps** The tensor_maps_native and tensor_maps_MNI152 folders include near-identical information, and contain native-space and MNI152-space maps of measures derived from the DWI data. Measures shared include fractional anisotropy (FA), mean diffusivity (MD), axial diffusivity (L1), and radial diffusivities (L2, L3). Example FA and MD maps in native space are shown in figure 1. To generate DWI metrics, we reconstructed tensor maps using FSL's DTIFIT[26,27]. Once constructed, the FA and MD maps of each participant were aligned with a multivariate registration using the ANTs toolbox[22,28] to the 'FMRIB158_FA_2mm' and 'FSL_HCP1065_MD_2mm_FMRIB58' respectively (FA and MD-based templates in MNI-152 space provided from FSL toolbox). This provided optimal alignment of both gray and white matter. Registration involved both linear (affine) and non-linear registrations (SYN-Diffeomorphic). Subsequently, linear and non-linear transformations generated from the prior registration were applied to all the tensor metrics using a trilinear interpolation. These methods are similar to those described previously[19].

**Connectomes** The 'derivatives/networks' folder contains structural brain networks (connectomes) generated using tractography and parcellated regions from the individuals' DWI and T1w MRI.

The parcellation used has been described by[29,30], and includes four different spatial scales (36, 60, 125, 250). Scale 36 is effectively the same as the widely used Desikan-Kiliany parcellation[31]. Other scales contain more fine-grained subdivisions of the previous scale, going from 82 regions at scale36 to 462 regions at the highest resolution (scale 250). These parcellated regions were included in our earlier data release (named 'myaparc_XX', where XX represents the scale), and have been used extensively in our previous work[7].

Tractography was performed using MRtrix3 software[32]; we first used 'dwi2response' with the 'dhollander' algorithm selected[33]. Once the response functions were calculated, we then estimated multi-tissue fiber orientation distributions (FOD) using the 'dwi2fod'[34] with 'msmt_csd'[35]. To benefit from anatomically constrained tractography (ACT), we make use of each subjects' corresponding FreeSurfer-segmented T1w image, inputting it into '5tt2gen' with 'freesurfer' and '-sgm_amyg_hipp' options enabled. Global intensity normalisation was preformed with 'mtnormalise', then gray and white matter segmentation from the output of '5tt2gen' was fed into '5tt2gmwmi'. Two different tractography algorithms were calculated; deterministic and probabilistic. Deterministic tracking used streamlines tractography based on spherical deconvolution (SD)[32]. Probabilistic tracking used the *i*FOD2 (Second-order integration over fiber orientation distributions) algorithm[36]. The output tracts of both methods were then filtered using the SIFT-2 algorithm with ACT enabled[37].

The following tracts and atlases were then input into 'tck2connectome'[38] for connectome construction. A connection between regions was identified as present if the streamline intersects the node at the endpoint. This resulted in the generation of connectivity matrices, where rows and columns represent different brain regions, and entries in the matrix define the connectivity strength.

There is no consensus on how to best define connection strength[39]. We therefore take an agnostic approach and share multiple networks per person, each with a different definition of 'strength'. We expect this offers the most flexibility to the interested researcher to choose the most appropriate measure relevant to their own research question. Overall, five different measures are included. Measure 1: Count reflects the total number of streamlines connecting two regions. Measure 2: CountScaled is a normalised measure of Count, scaling streamline count on the inverse of streamline length and size of two node volumes[29]. Measure 3: MeanFA measures the mean FA along the streamlines, while Measure 4: MeanMD is similar. Measure 5: MeanLength should not be interpreted as strength but measures the mean length of connecting streamlines in millimeters.

In addition, tractography can be prone to early failure in proximity to the grey/white matter boundary due to partial volume effects. We therefore also release a version with grey matter regions dilated by 1mm into the white matter using previously described dilation techniques[19].

In total, we therefore release 80 connectivity matrices per person (2 tractography methods [deterministic and probabilistic] x 4 parcellations [scales 36, 60, 125, 250] x 5 'strength' measures x 2 dilated or not).

## Quality control

All unprocessed and processed images were viewed using interactive matrix viewer 'imshow' using the packages: plotly (6.3.0)[40] and matplotlib (3.9.4)[41] in Python (3.9.6). We used the ANTs python toolbox[42] to import all NIFTI images into a numpy (1.26.4)[43] matrix format. Five spaced out coronal slices of each subject were stacked to their respective total group of coronal slices for all subjects. Then, using 'imshow' and 'write_html' in plotly, we viewed each of the five interactive plots of all subject images.

As connectome parcellations were generated per individual FreeSurfer parcellation, we decided to check all corresponding rows and columns of all matrices, ensuring that each row/column represented the same region for all subjects. Therefore we analysed the index number of each freesurfer region to the atlas NIFTI image using the ANTs python toolbox, and matched it to the index number in the row and column of the index. Match checking involved using the tool '.isin()' in the Pandas toolbox (2.3.1)[44] in python. Subjects with one or more missing regions in their freesurfer atlas were removed. In addition, z-scores for all connectome matrices were calculated per parcellation scheme and measure. Then z-score matrices were flattened into a one dimensional space and stacked per subject to create an overall 2D matrix display of z-scores (Connections x Subject). Then interactive viewable z-score matrices for each parcellation scheme and measure was generated and checked using 'imshow' and 'write_html' in plotly. Subjects with strong ± z-scores had all scans and matrices re-checked and either removed or reprocessed.

## ENIGMA study replication

As a demonstration of this public datataset's potential, we sought to replicate findings in a landmark previous study[11]. As such, we re-ran the ENIGMA TBSS pipeline using tract based spatial statistics (TBSS) applied to the processed MNI152 FA maps.

Ultimately, the TBSS and ENIGMA pipeline processing resulted in a spreadsheet containing FA values across subjects for each WM region of interest. This spreadsheet is shared in the data release.

## Statistical analysis

We computed abnormalities in each white matter (WM) region of interest (ROI) for each subject, similar to our previous approach for grey matter[7]. We controlled for biological and technical covariates known to affect the signal[45]. As in[11], we first harmonized FA values across acquisition protocols using ComBat[46]. We then fitted a generalized additive model (GAM) to the healthy control data to account for age and sex effects and used the fitted model to compute residuals for all subjects. Abnormalities were quantified by z-scoring each subject's residuals relative to the mean and standard deviation of control residuals.

When calculating abnormalities for a control subject, that subject was excluded, and the remaining controls defined the reference distribution. Finally, we estimated effect sizes between subjects and controls using Cohen's $d$ applied to the z-scores. Statistical significance was estimated using Wilcoxon rank-sum tests.

We also compared IDEAS II white matter effect sizes with those from the original ENIGMA study using Pearson correlation. Effect sizes for the ENIGMA study were extracted from table S4 in[11]. It should be noted that of the 1249 patients and 1069 controls, in the ENIGMA study, 51 patients and 29 controls overlap with the IDEAS cohort representing a potential minor confound.

## Abnormal cluster detection

As a demonstrative research example, we apply our abnormal cluster detection algorithm[47]. In brief, the algorithm computes abnormalities in each voxel, by comparing against the mean and standard deviation observed in healthy controls. Next, voxel abnormalities are thresholded and clusters of abnormalities are identified. Further details can be found in[47].

## Data organization

Raw, and preprocessed scans are shared in an identical structure using The Brain Imaging Data Structure (BIDS). Given that multiple sessions are important in this release, we have included session information (denoted 'ses-1' or 'ses-2'). Supplementary figure S2 summarizes the data layout.

## Data availability

All data is shared openly and can be accessed at the locations in Table S1. Data will also be linked publicly at www.cnnp-lab/ideas-data, and openneuro.org upon acceptance of the manuscript.

# Results

## Reduced Fractional anisotropy in Temporal Lobe Epilepsy

A landmark study investigated alterations in FA across epilepsy syndromes across a large cohort of data from 21 sites[11]. In that study the authors used tract-based spatial statistics to identify changes in major white matter tracts. In figure 2 we sought to replicate the analysis in left and right TLE (panels A and B respectively). The most pronounced changes in our data are located in the external capsule (EC), uncinate fasciculus (UNC), and parahippocampal cingulum (CGH; also known as the ventral cingulum) ipsilaterally. The average FA was also substantially reduced in the individuals with epilepsy.

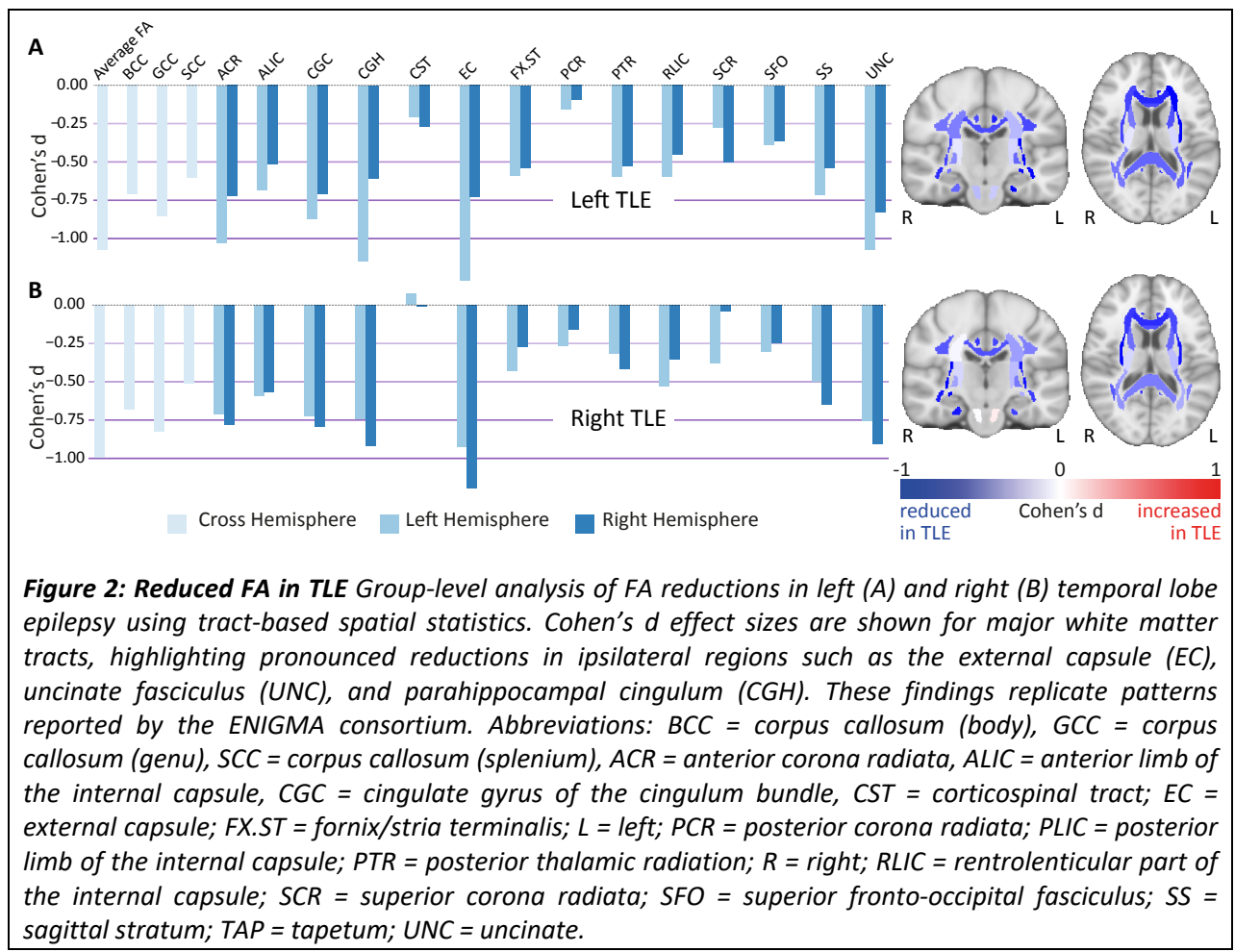

*Figure 2: Reduced FA in TLE* Group-level analysis of FA reductions in left (A) and right (B) temporal lobe epilepsy using tract-based spatial statistics. Cohen's d effect sizes are shown for major white matter tracts, highlighting pronounced reductions in ipsilateral regions such as the external capsule (EC), uncinate fasciculus (UNC), and parahippocampal cingulum (CGH). These findings replicate patterns reported by the ENIGMA consortium. Abbreviations: BCC = corpus callosum (body), GCC = corpus callosum (genu), SCC = corpus callosum (splenium), ACR = anterior corona radiata, ALIC = anterior limb of the internal capsule, CGC = cingulate gyrus of the cingulum bundle, CST = corticospinal tract; EC = external capsule; FX.ST = fornix/stria terminalis; L = left; PCR = posterior corona radiata; PLIC = posterior limb of the internal capsule; PTR = posterior thalamic radiation; R = right; RLIC = rentrolenticular part of the internal capsule; SCR = superior corona radiata; SFO = superior fronto-occipital fasciculus; SS = sagittal stratum; TAP = tapetum; UNC = uncinate.

To quantitatively investigate our findings' similarity with those described by[11], we performed a correlational analysis of the effect size. In figure 3 each data point represents a single white matter region, with y-axis values as shown in figure 2. X-axis values are effect sizes reported by[11] in their supplementary table 4. A high correlation is noted in both left, and right TLE (Pearson's rho=0.92 and 0.71 respectively).

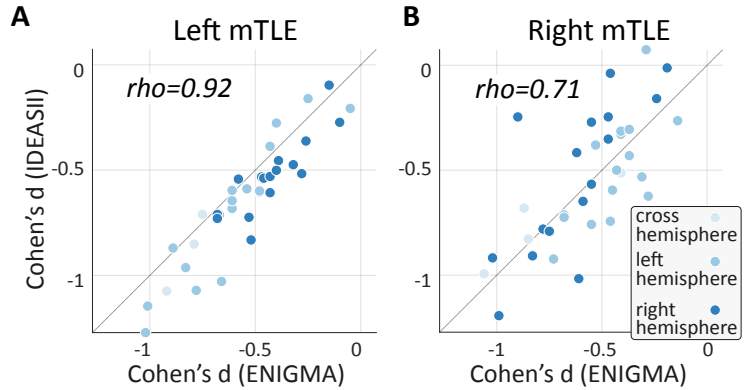

*Figure 3: Similar effects in IDEASII and ENIGMA. Scatter plots comparing FA effect sizes from IDEASII (y-axis) with those reported by ENIGMA (x-axis) for left (A) and right (B) temporal lobe epilepsy. Each point represents a white matter tract. Strong correlations (rho = 0.92 for left TLE; rho = 0.71 for right TLE) confirm consistency between IDEASII and ENIGMA findings.*

## Diffusion metrics to identify abnormality in individual patients

Next, we demonstrate the potential for DWI metrics to hold localisation information in an individual patient (figure 4a,b) and in the cohort (figure 5). To this end, we applied an abnormality clustering technique described previously[16]. The technique includes z-scoring of individual voxels' MD measures, by comparing to a control distribution after spatial alignment. Our release of MNI space MD maps facilitates this analysis. Following z-scoring, a clustering approach is used to identify voxels beyond a threshold that are spatially contiguous.

Figure 4a shows an individual patient to which the methods have been applied and shows a clear abnormality of increased MD in the white matter of the right temporal lobe. This patient had a subsequent right anterior temporal lobe resection (figure 4b) and was seizure free postoperatively.

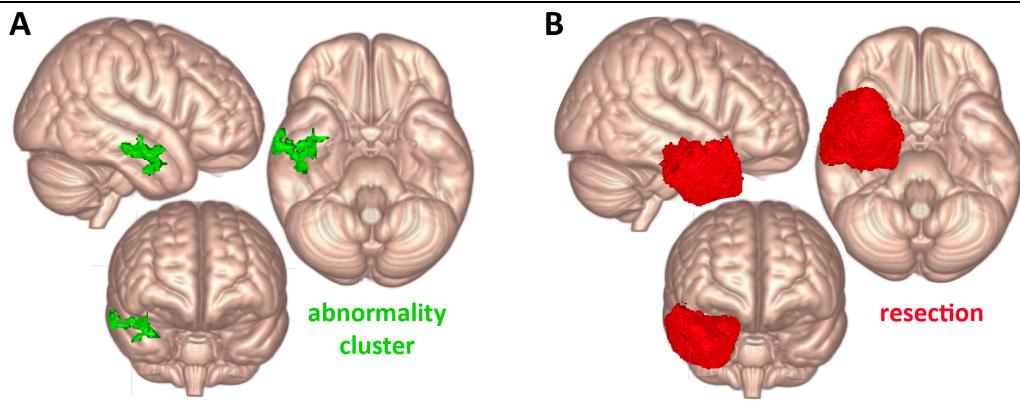

*Figure 4: dMRI abnormality in a single patient. Example patient analysis using voxel-wise abnormality clustering based on MD maps. (A) Abnormality cluster of increased MD identified in the right temporal lobe; (B) corresponding surgical resection mask.*

Across the cohort, 102 patients had anterior temporal lobe resections and were free of disabling seizures 12 months postoperatively (ILAE1 or 2). Of these patients, figure 5 shows the locations of their abnormality clusters. The most common locations by far are in the ipsilateral temporal lobe in the hippocampus and temporal white matter (indicated by red arrows in figure 5).

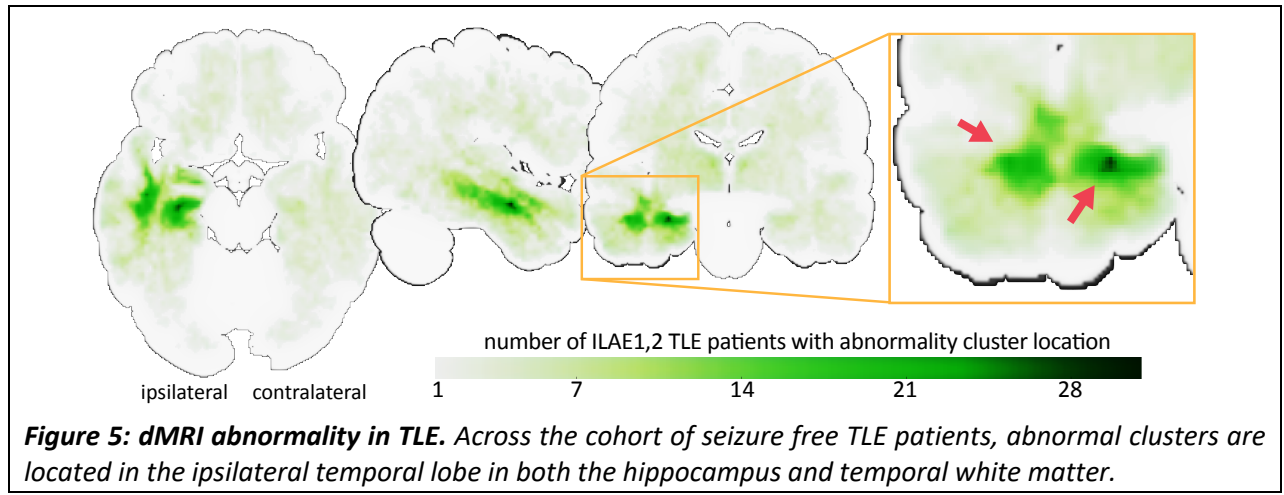

*Figure 5: dMRI abnormality in TLE.* Across the cohort of seizure free TLE patients, abnormal clusters are located in the ipsilateral temporal lobe in both the hippocampus and temporal white matter.

## Streamline connectivity amenable to network analysis

In addition to being sensitive to the abnormalities shown in figure 2, a major advantage of DWI is that it allows the inference of white matter connectivity. An example subject is shown in figure 6 to demonstrate this, and is the same patient shown in figure 1. Based on our first data release, parcellated regions of interest can be seen and used to define brain network 'nodes' (figure 6a). Figure 6b presents an advance with the new IDEASII data release; the DWI derived streamline tractography. Here we show only the streamlines with both ends terminating in resected regions (figure 6c). This allows the more abstract visualisation of the network shown in figure 6d, where spheres represent the centroid of each brain region (network node) and lines represent the streamline count between regions. Such formalisation of the brain as a network allows for further analysis of e.g. associations with seizure spread[48] and lateralisation[49].

We openly share networks for 191 patients and 97 healthy controls. Networks shared are whole brain using multiple parcellation options. When used in conjunction with resection information, subnetworks can be investigated which approximate the surgically removed subnetwork. Such (sub)networks are amenable to further connectomic and graph theoretic analysis[50], or virtual brain modelling[51].

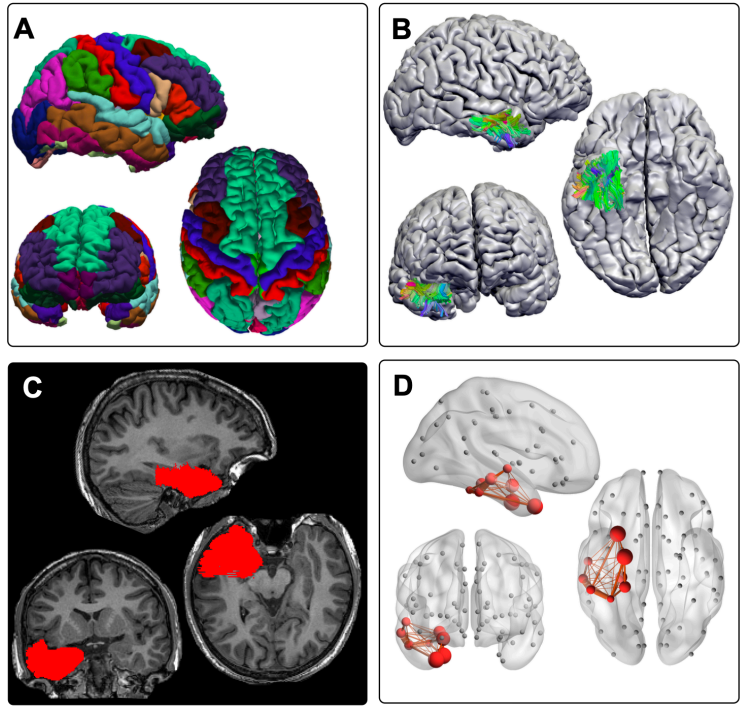

*Figure 6: Streamline tractography and network representation for an example patient.* *(A) Parcellated cortical regions defining network nodes; (B) tractography involving the resected nodes (C) resection mask (D) abstract network visualization showing nodes (brain regions) and edges (streamline counts) constrained to only those which begin and end in resected regions.*

## Discussion

We compiled, processed, and analyzed a large diffusion-weighted MRI dataset comprising 216 individuals with epilepsy and 98 healthy controls. Our analysis revealed widespread alterations in fractional anisotropy across multiple white matter regions, most pronounced in the ipsilateral hemisphere, consistent with previous findings from the ENIGMA consortium[11]. Additionally, we demonstrated potential clinical utility with abnormality localisation, and streamline tractography in a patient with right temporal lobe epilepsy. Finally, in line with our commitment to open science, all data and networks are made available through widely accessible platforms.

In recent decades there has been a shift in perspective to consider epilepsy as a network disorder[52,53]. Multiple studies have shown various brain network properties are related to epilepsy at diagnosis[54], seizure spread[48], and treatment response, in both adults[55] and children[56]. Although the shift towards brain networks in epilepsy has been popular, it should be noted that clinical localisation of epileptogenic tissue is still predominantly based on a 'focal' perspective. In reality there are likely multiple mechanisms leading to seizures forming a spectrum of focal to network abnormality, potentially with central thalamic drivers[57]. The data shared here may be of use to investigate this spectrum from focal regions to wider networks.

The heterogeneity of epilepsy (in lesion location and potential underlying mechanism) makes it challenging to study. This typically leads researchers to study subgroups such as TLE in isolation of other epilepsies. To mitigate issues around heterogeneity, one approach is to use normative modelling approaches[58]. One study has already made use of the original IDEAS data release by using normative modelling to identify altered cortical thickness in TLE[59,60]. For such approaches to work effectively it is crucial to ensure healthy control scans are available to perform batch correction. By releasing control data, additional to that from patients, we expect much wider use, and more robust findings[60,61].

Although diffusion-weighted MRI has traditionally been acquired to identify critical white matter pathways for surgical avoidance, emerging evidence suggests it may also aid in seizure localization. Figure 4 supports this view and replicates findings from our previous work with this data[47]. These observations highlight the promise of quantitative, objective neuroimaging in guiding surgical decisions. Realising this potential, however, requires large-scale datasets such as IDEAS to enable robust algorithm development. Such algorithms may be statistical, as demonstrated here, or leverage deep learning and AI approaches, which demand extensive training data.

Here, we investigated the two most commonly analysed DWI metrics, namely FA and MD computed from the diffusion tensor model. A more complex model may be beneficial with higher sensitivity and easier interpretation[62,63]. To compute measures from more complex models a more advanced acquisition protocol is typically required. This release of IDEAS contains 98 patients and 69 control scans with multi-shell protocols. This allows the opportunity to investigate multi-compartment models such as NODDI[64]. Although not analysed here this offers exciting potential directions for future research.

Some previous studies from our group analysed some or all of the data in the IDEAS releases. Here we summarise a selection of those works.[48] investigated network associations with focal to bilateral tonic-clonic seizures (FBTCS), and found greater and more widespread network abnormalities in people with FBTCS than those without.[65] related abnormality in major limbic tracts with epilepsy duration, reporting a heterogeneous effect.[19] compared white matter at different spatial depths in the temporal lobe, suggesting that superficial white matter may be more sensitive than deep white matter areas for lateralisation in TLE. In a multimodal analyses[16] compared DWI and intracranial EEG abnormalites, and[66] compared DWI and T1w abnormalites for localisation and outcome prediction. The most recent study from our group has combined DWI, T1w MRI, and intracranial EEG abnormality, achieving outcome classification of AUC=0.92, highlighting the importance of multimodal data integration[kozma2025b?].

We envisage several potential research uses of these data. Artificial intelligence algorithms are becoming more common, with pre-trained models now available for prediction of demographic factors, and ability to improve image quality or perform other automated analyses[67,68]. Given the clear labels of resected regions available from our first release, and demographic/clinical information, we expect the IDEAS data to be of use for training and testing of AI models. A second area in which we anticipate these data could be useful is for network neuroscience (connectomics), with potential for use of modern network statistical approaches[69,70]. A third use case is the area of virtual brain modelling. Virtual brain models typically comprise a set of equations to simulate brain dynamics, with connectivity parameters inferred from patient data[51]. By modifying model parameters it is possible to simulate treatments such as surgery or stimulation. Our pre-processed connectivity matrices can be directly imported into such models, facilitating wider use. Other, more traditional neuroimaging analyses are also possible.

Several limitations should be acknowledged. First, all data were acquired at a single site, which may constrain generalizability across patients. Second, the dataset includes two acquisition protocols, which could introduce subtle differences in diffusion metrics despite preprocessing and harmonization efforts, but may aid generalizability across protocols. Third, the small overlap with the ENIGMA cohort may represent a minor confound in our replication analyses.

A major strength of the IDEAS dataset lies in its integration of multiple imaging modalities with resection masks and rich clinical and demographic information. Importantly, all data presented here are in addition to previously released resources. This means that for many individuals, the following information is available: pre-operative imaging (including T1-weighted MRI, FLAIR MRI, and diffusion-weighted MRI), a resection mask identifying tissue subsequently removed, as well as age, sex, age at onset, medication history, and annual outcomes for up to five years post-surgery. The depth and organization of this dataset, shared at such scale, is unprecedented. We hope that its availability will accelerate new discoveries and improve outcomes for individuals living with epilepsy.

# Acknowledgements

We thank members of the Computational Neurology, Neuroscience & Psychiatry Lab (www.cnnp-lab.com) for discussions on the analysis and manuscript; P.N.T. and Y.W. are both supported by UKRI Future Leaders Fellowships (MR/T04294X/1, MR/V026569/1, MR/Y034104/1). JSD, JdT are supported by the NIHR UCLH/UCL Biomedical Research Centre.The authors acknowledge the facilities and scientific and technical assistance of the National Imaging Facility, a National Collaborative Research Infrastructure Strategy (NCRIS) capability, at the Centre for Microscopy, Characterisation, and Analysis, the University of Western Australia. GPW acknowledges support from MRC (G0802012, and MR/M00841X/1).